\documentclass{elsart}
\usepackage{graphicx}
\usepackage{subfigure}
\begin{document}
\begin{frontmatter}
\journal{arXiv. org}
\title{Magnetism, entropy, and the first nano-machines}
\author{Gargi Mitra--Delmotte\thanksref{gmail}}
and
\author{Asoke N Mitra\thanksref{mmail}}
\thanks[gmail]{Corresponding author, present address: 39 Cite de l'Ocean, Montgaillard, St.Denis 97400, REUNION;  e.mail: gargijj@orange.fr ;
Tel. and Fax. no.: 00-262-262307972 }
\thanks[mmail]{Formerly INSA Einstein Professor, Department of Physics, Delhi University; address : 244 Tagore Park, Delhi -110009. INDIA ;
e.mail: ganmitra@nde.vsnl.net.in }
\date{}
\maketitle

\begin{abstract}
The efficiency of bio-molecular motors stems from reversible
interactions $\sim$ $k_B T$; weak bonds stabilizing intermediate
states (enabling $direct$ conversion of chemical into mechanical
energy). For their (unknown) origins, we suggest that a magnetically
structured phase (MSP) formed via accretion of super-paramagnetic
particles (S-PPs)  during serpentinization (including magnetite
formation) of igneous rocks comprising the Hadean Ocean floor, had
hosted motor-like diffusion of ligand-bound S-PPs through its
'template'-layers. Ramifications range from optical activity to
quantum coherence. A gentle flux gradient offers both
detailed-balance breaking non-equilibrium and $asymmetry$ to a
magnetic dipole, undergoing infinitesimal spin-alignment changes.
Periodic perturbation of this background by local H-fields of
template-partners can lead to periodic high and low-template
affinity states, due to the dipole's magnetic degree of freedom. An
accompanying magnetocaloric effect allows interchange between
system-entropy and bath temperature. We speculate on a magnetic
reproducer in a setting close to the submarine hydrothermal
mound-scenario of Russell and coworkers that could evolve
bio-ratchets.
\\


\end{abstract}

\end{frontmatter}

\newpage

\section* {1.1 Introduction: Bio-molecular dynamics}
Increasingly it is becoming apparent that the dynamics in biology at
the nanoscale, such as in molecular motors, is part of the generic
phenomena governed by the (linear) fluctuation-dissipation theorem
[1,2]. The diffusive movement of the system traversing between two
energy states, aided by random Brownian forces, gets rectified by
coupling to a non-equilibrium force, e.g. ATP hydrolysis [3].  And
such systems with low Reynold number, access to internal degrees of
freedom, plus asymmetric interactions, can thrive on {\it the best
of both worlds}-an equilibrium local state that can harness thermal
fluctuations in a diffusive step as well as the directionality
governed by a non-equilibrium reaction -- leading to net movement in
an asymmetric yet periodic energy landscape [4]. This brings to
light two essential requirements for executing such dynamics:
Firstly, weak bonds (hydrophobic, hydrophilic, van der Waal's,
H-bonds, etc) help to 'pin-up' the motor temporarily in different
equilibrium states. Secondly, the continuous nature of the energy
landscape connecting different states shows a system that can absorb
energy in an essentially continuous and reversible manner
(adiabatic) by utilizing energy from random Brownian hits ($\sim$
k$_B$T) which is of the same scale as rotational energy states in a
molecule. This feature similarly enables a physical nanosystem to
undergo periodic cycles a la Berry's phase, between the two states
[5].

\section*{1.2 Origins of weak bonds and reversible interactions}
So the question arises: How could such reversible interactions and
intermediate states, underlying the efficiency of these machines,
have been physically realized by matter present at the dawn of Life?
Note that these are almost impossible to achieve using chemical
bonds that are the very basis of proposals for template-based
processes using mineral crystal surfaces. Traditionally,
origin-of-life theories concentrate either on its replication or
metabolism aspects. On the other hand, the origin of the ubiquitous
molecular motors, in and across all living systems, is seen as a
later addition. According to Vale [6] two inventions were important
in the development of motors: one-dimensional electrostatic sliding
along polymers, and a conformational-change mechanism in the active
site of a nucleotidase enzyme. How this happened however has been
left unaddressed and is largely unknown. Here we suggest that Life's
origin was strongly linked to the emergence of nano-systems
utilizing the thermal energy of the surroundings, just as in today's
biological nano-machines. A physical phase captured before the onset
of the mineral crystallization could have also hosted template
processes proposed by Cairns-Smith [7]. This is line with Dyson's
[8] proposal for `physical reproduction' (not 'chemical
replication') as having initiated Life, in a metabolically enriched
environment. In fact, $physical$ forces seem to be well equipped to
deal with some of the logical difficulties cropping up with
'chemistry-only' origin-of-life approaches. Indeed, the power of
$magnetic (H)$ fields for external control on super-paramagnetic
matter, seems to have gone unnoticed hitherto, despite their
omnipresence in space and time. To begin with, H-field controlled
super-paramagnetic particles (S-PPs) could have provided a ready
basis at the origins of life for generating the various
energy-transduction systems coupling the formation/use of
energy-rich molecules with temperature or charge (redox, pH )
transferring gradients [9]. Only later could the non-equilibrium and
symmetry-breaking aspects of the field have been replaced by
energy-rich molecules and asymmetric interactions. Besides, the long
appreciated spin-system mimicking features across myriad phenomena
displayed by biological soft-matter (landscape processes,
orientational order in fluid state, etc.), would be easier to
understand $logically$ if we had begun with a $magnetic$ Ancestor in
the first place. After all, many bacteria play host to magnetosomes
of of greigite $(Fe_3S_4)$ [10, 11]. (And, biomineralization using
magnetite ($Fe_3O_4)$ --a close relative-- is considered the most
ancient matrix-mediated system that could even have served as an
ancestral template for exaptation [12]). Sure enough, magnetic
alignment of a particle to its partner in a template (an array of
aligned particles) could have provided the beginnings for embodying
weak bonds, typical in biology. Here, local reversibility at each
infinitesimal step is achieved via effectively continuous spin
alignment changes, where maximum/minimum interactions lead to
association/dissociation, respectively. These could be driven by
simple thermal fluctuations, just as in today's bio-molecular
motors.

\section*{1.3 Need for a dynamical lattice: Power of magnetism}
Now, field-induced structure formation as seen in non-ideal magnetic
fluids (Sect.2.1), could bring about orientational long range order
in the aqueous dispersed particles. Such a dynamic array as the
'fountainhead of Life' represents a major departure from
conventional template approaches based on either crystal-surfaces,
or where evolved chemicals had sufficient complexity for spatially
asymmetric interactions for self-assembly (e.g. liquid crystals).
However, no convincing explanation seems forthcoming as to how a
reproducing life-like assembly from complex molecules had evolved
from an immense medley of compounds. Consider instead, the role of
magnetic dipolar interactions in giving rise to a dynamic assembly,
without having to wait for the evolution of complex molecules, whose
feasibility is crucial for the emergence in the Hadean of molecular
motors -- key players across kingdoms in biology. While a similar
passage would be impossible with rigid lattices of mineral crystals,
the simpler possibility of physically reproducing [8] magnetic
particles exists for extending the horizons of traditional
approaches by combining chemistry with myriad physical effects. In
this scenario, chemistry continues to play a role in the ligand
shell reactions of colloidal S-PPs as in the simulations of
Milner-White and Russell [13], but magnetic accretion provides the
$confining$ force for herding them together. Cutting through
dipole-dipole interactions holding together layers of a magnetically
ordered phase would require energy, orders of magnitude less than
those cementing crystal layers. At the same time an
orientation-based magnetically herded 'array' would retain the
information transmission feature of ordered crystal surfaces [7].
Again, not only crystal surfaces, but individual layers of a dynamic
array could be imagined as the very templates a la Cairns-Smith that
hosted transfer reactions in the origins of Life. The tremendous
increase in surface area vis-a-vis mineral crystal surfaces would
also similarly stretch the prospects of catalytic activity, crucial
for metabolism, and in line with today's spotlight on the nanoscale.
Indeed, $packing$ in physically accreted finite systems comes ready
with built-in aperiodicity, as an effective substitute for the
superimposed aperiodic distribution of metal ions on infinite
periodic crystal lattices [14]. This very feature underlies the
efficient packaging of information in nucleic acids, where the lack
of correlations across sequences (random nature) satisfies Claude
Shannon's maximum entropy requirement [cf. 15].

\section*{1.4 Outline of paper}
With this background, we shall first briefly review magnetically
structured phases (MSPs) of dispersed magnetic colloids (Sect.2.1),
for trying to identify the ingredients required for extending this
scenario to a possible Early Earth (Hadean) Ocean Floor setting
(Sect.2.2). Next in Sects.3.1-3.5, we present a detailed
correspondence (mapping) of the features of bio-molecular motors
with those of super-paramagnetic particles diffusing through a
magnetically structured phase. Finally, in Sects 4.1-4.2, we ask how
bio-ratchets could have originated and suggest a greigite-based
scenario; Sect.4.3 concludes with a discussion on the potential of
magnetism.

\section*{2.1 Field-induced aggregates in non-ideal ferrofluids}
The above brings us to the well known area of ferrofluids: colloidal
single-domain magnetic nanoparticles ($\sim$ 10nm) in non-magnetic
liquids that can be controlled by moderate H-fields ($\sim$ tens of
milliTesla) [16]. Coatings stabilize these dilute dispersions
displaying ideal single-phase behaviour due to prohibited (chemical)
inter-particle contacts. In contrast, the present application
concerns the interactions within the magnetic subsystem, while the
carrier remains in the liquid state. The deviation from ideal
magnetization behaviour shows up on increasing particle
concentrations that can be understood in terms of H-field-induced
inter-particle interactions leading to internal structure formation
and manifesting in {\it dense phases} -a milder phase transition
than to the solid-crystalline one. The structure of hydrated,
heterogenous aggregates (e.g. chain-like, drop-like, worm-like
micelles) would depend on factors like the strength of the applied
field, the nature of the ferrofluid (molecular shape,
susceptibility, etc.) [16, 17, 18]. An increase of chain size beyond
a critical length, compactification due to interparticle magnetic
interactions, formation of globules as nuclei for new dense phase,
are all seen in the scheme of phase transitions leading to formation
of bulk drop-like aggregates. As to the role of polydispersity, Wang
and Holm [19] found that the fraction of large particles, with
larger relative dipole moments in proportion to their volume, would
overcome thermal forces more easily and respond to weaker fields and
therefore dictate magnetization properties (e.g. initial
susceptibility), even in case of dilute fluids. Furthermore, the
solvent could also affect the aggregation, for Taketomi et al [20]
observed field-induced macrocluster formation in water and
paraffin-based ferrofluids but not in an alkyl-napthalene based one
(even at 0.2 Tesla). In contrast, macroclusters formed in the
water-based fluid at very low fields and remained even after
removing the field. Li et al [21] have pointed out the dissipative
nature of the field-induced aggregates that break up in response to
thermal effects upon removal of field. They propose a gas-like
compression model - a phase transition in which a particle
concentrated phase separates from a dilute one, by following the
orientation of the particle moments in the direction of the field.
And, the higher the field intensity the more compact the aggregates;
so that the aggregate space containing particles would decrease,
just as in a compressed gas. In this model, the total magnetic
energy of ferrofluids obtained from an applied field: $W_T = W_M +
W_S $; where $ W_M = \mu_{0} M H V $ and $W_S = -T \Delta S $ are
the magnetized and the structurized energies, respectively, $V$ is
the volume of the ferrofluid sample and $ \Delta S $ is the entropic
change due to the microstructure transition of the ferrofluid. An
assumed equivalence of $W_T$ (zero interparticle interactions), with
the Langevin magnetized energy $W_L = \mu_{0} M H V $ necessitates
to a correction in the magnetization, in terms of the entropy
change. Evidently, these systems are well equipped to analyze the
interplay between competing factors -dipolar interactions, thermal
motion, screening effects, etc. leading to the emergence of MSPs
[22]. Their colloidal state and magnetic entropy property can
provide a ready basis for mapping with complex biological
soft-matter. We now take a closer look at bio-molecular motors, as
these systems capture many of the complexities of biosystems.

\section*{2.2 Magnetic assembly on the Ocean floor}
Analogous to non-ideal ferrofluids with interparticle interactions,
three ingredients are required for a dynamic lattice: (1) the
presence of a moderate local H-field on the Hadean Ocean Floor; (2)
a newly forming super-paramagnetic suspension turning into tiny
magnets due to no. (1); and (3) $charge$ on particles. Serpentinized
and magnetized igneous rocks [23, 24] could offer a local field (see
Sect.4.2), with geochemistry providing the rest (Sect.4.2). Note
that in contrast to homogeneous synthetic ferrofluids, a suspension
forming {\it in situ} in presence of rocks, would have likely been
polydisperse, with larger particles dictating magnetization
behaviour (Sect.2.1).

\par
Simulations of field-induced dissipative structures in non-ideal
ferrofluids (see above) postulate the energy of a constituent
particle to have contributions from the dipole-dipole interactions
with neighbours; repulsive (charge/steric) effects; and its energy
accruing from its orientation w.r.t. the H-field [25]. In the
absence of a complete theory of dipolar fluids, and on the basis of
available literature [25, 26], we envisage the emergence of an MSP
upon gradual build-up of particles interacting via dipole-dipole
interaction, just above the rocks. This, in turn would increase
magneto-viscosity, impeding particles from rotating freely.
Contributing factors for dipole ordering with energy-minimization
include material properties, ligand-field effects, polydispersity,
H-field strength, apart from ordering-variation [27] from parallel
to anti-parallel. And, transient chains/arrays forming due to
interacting dipoles forming layers of the MSP, could serve as
magnetic templates for enabling bio-molecular motor-like transport.
Figure 1 represents roughly parallel orientational correlations with
resultant magnetization of MSP along the rock H-field. Finally,
requirement no. (3) (charge on particle) is chosen in view of the
key role of conflicting forces - here attractive magnetic, and
repulsive electrostatic, -- in bringing about a dynamic assembly
[see 28]. Again, in the absence of `synthetic coatings', the high
ionic strength (screening effect) of the sea water [29] would have
further encouraged dipole-dipole interactions.

It is interesting to compare a corresponding build-up of particles
under ambient temperature, in the absence of a magnetic field. These
would gradually form a colloidal network which would be expected to
$age$ by passing on to the crystalline phase. Thus it seems that a
moderate local H-field $pre-empts$ this process by capturing the
build-up of magnetically tunable particles, thereby enabling
Life-like dynamics in an otherwise inaccessible magnetically ordered
fluid phase. Note that unlike a chemically bonded thermally formed
gel, a magnetic gel would retain the potential of reverting back to
its colloidal components just like colloid-gel transitions pointed
out in living systems [30].

\section*{3.1 Directed motor movement: questions of origins}
Thus far, we have considered the possibility of a magnetically
ordered phase on the Hadean Ocean floor. But, how could directed
diffusion as in bio-molecular motors migrating on templates like
proteins and nucleic acids, have occurred for particles diffusing
through these MSPs? In brief, motor proteins normally display
unidirected transport, walking towards either the plus (e.g.
kinesins) or the minus (e.g. dyneins) ends of the template (e.g.
filaments, microtubules) that are polar polymers, arranged in a
head-to-tail fashion. With cargo bound to their tail end, the motor
diffuses back and forth till its capture by sites ahead in the
progress direction; the greater likelihood of which follows
asymmetry in binding affinity. The key change due to the ATP ligand
is thus an altered energy landscape potential, leading to states
with altered binding affinity. Although the free energy of this
'bound' conformation is larger than the minimum free energy of the
ligand-free protein, thermal motion makes this conformation
accessible [31].  Recent experiments by Taniguchi et al [32], led
them to propose an entropic basis of rectification for the directed
migration of kinesin; the backward step leads to a significantly
lower entropic state than in the forward one.

Two outstanding clues are thus retrieved from motor dynamics: First
is the capacity of the motor to combine with local anisotropy to
bring about net movement which stems from its {\it internal degrees
of freedom} allowing it to take on a different trajectory (different
intermediate states on altered landscape potential) for the second
half-cycle in each period. Second is its apparent capacity to
undergo infinitesimal conformational changes by extracting energy
$\sim k_B T$ from the thermal bath with help from
close-to-equilibrium coupling to a non-equilibrium source for
rectifying these fluctuations. Then in the Hadean, in the absence of
complexity, we encounter the following question: How about particles
bearing internal degrees of freedom having carried out similar
entropy-reducing {\it Maxwell Demon-like} feats, as occurs for small
systems [1], in the origins of Life? To address this question, we
shall extrapolate this scenario to the directed diffusion of S-PPs
through layers of an MSP, and check if the latter can supply both a
topologically and energetically satisfying correspondence to the
rectified diffusion of molecular motors on templates.

\section*{3.2 Directed diffusion of dipole through MSP}
Recall that a magnetic particle moves in response to a
field-gradient. A uniform field can orient a magnetic dipole but as
the forces on its north and south poles would be balanced, there
would be a zero net translational force acting on it. This situation
would change in the presence of a field gradient. And, magnetic
field lines due to rocks would be cutting through the MSP. Their
nature would be expected to be non-homogeneous, albeit changing in
intensity in a very gradual manner. If the variations were so gentle
as to appear small as compared to the radius of the diffusing S-PP,
it would sense an effectively isotropic local environment [33,34].
The {\it symmetry-breaking effect} of the gradient would be felt by
the particle at greater distances and {\it bias the directional
preference for diffusion}. This diffusion of the nano-particle
(negligible inertial effects) would further slow down to a net drift
if it were to take place in a magneto-viscous medium formed as a
result of magnetic dipolar forces between S-PPs. Next, two changes
are expected upon ligand binding: lowering of both rotational
freedom and coercivity [35] on ligand-bound end. Thus, while
unconstrained rotation of ligand-free particles enables alignment
and propagation of the `information' in the magnetic dipole-ordered
assembly ('reproduction'), ligand-binding aids diffusive passage.

Further, diffusion is expected to be faster for particles with
increased magnetization. Although this would also depend on the
nature of ordering in particle clusters, our choice of criterion no.
(3) ensures that only small size particles would diffuse through the
MSP. A charge on the S-PPs would have not only aided in
self-organization of the structured magnetic phase, but also its
layers of similarly charged particles, would have repelled the entry
of large clusters of similar charge-carrying particles. This effect
was likely accentuated due to the low effective shielding (low
concentration of sea-water) inside the layers of the dense MSP. And
while mono- or di-mers could be expected to diffuse through, higher
molecular weight members with increased surface charge would face
resistance to passage, due to greater repulsive effects. (The
oriented diffusion of S-PPs (see Figure 1) is imagined in the
direction parallel to that of the field lines, so $\sin \theta  = 0$
and therefore no force is exerted by the H-field on a charge moving
parallel to it). In figure 1, the S-PPs (in blue) have both motional
and spin degrees of freedom, in contrast to their MSP-counterparts
embedded in a magnetically bonded network with orientational
correlations (in black).

\begin{figure}
\centering
\includegraphics[width=120mm,height=100mm]{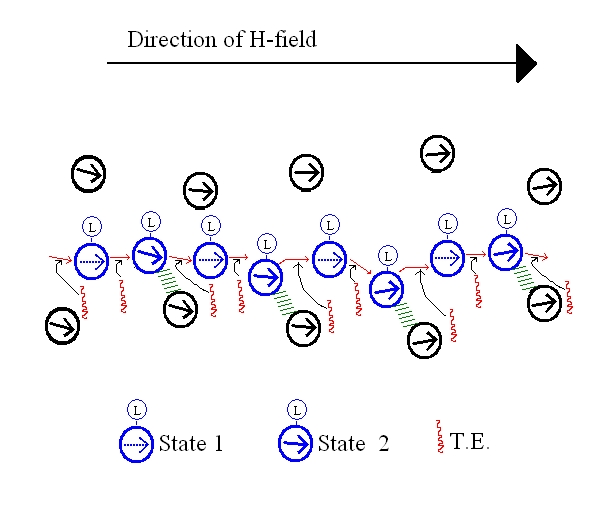}
\caption{Directed interactive diffusion of S-PP through MSP (with
parallel correlations). MSP represented in black;State 1/ State 2:
lower/higher template-affinity states of the ligand (L) -bound S-PP,
in blue; green lines signify alignment in State 2; T.E. or thermal
energy from bath; rock H-field direction indicated on top of figure,
see text}
\end{figure}

\section*{3.3 Interactive cycles}
The periodically changing landscape potential of complex
bio-molecular motors seems to match the periodic perturbations on
the background rock H-field due to superimposed local H-fields of
particles, constituting the MSP layers (the `templates'), as `seen'
by the S-PPs drifting in the gentle gradient. This is caused by the
variations in orientations of individual template-partners, even
while the resultant MSP-magnetization remains along the rock H-field
direction. (For, the local magnetic field acting on a particle is
the sum of the external field and the dipolar fields of the other
particles [26]). Thus, spin-ordering in a diffusing S-PP, oriented
along the rock H-field - State 1 having lower template-affinity --
would change for aligning to the local H-field of a template
partner--State 2 having higher template-affinity. These changes
would be similarly facilitated by thermal excitations from bath
[c.f. 36, 37], with rectification by either the gentle H-field
gradient or local template-partner H-fields. Indeed, this scenario
of changing H-fields for modulating intrinsic dipole-dipole
interactions closely resembles the simulations by the Korenivski
group [38, 39] who propose a ferrofluid -based associative neural
network for pattern storage where the respective transition
probabilities satisfy detailed balance. These demonstrate how local
variations of the external H-field (via Zeeman effect) can be used
to influence the positions and spin orientations of individual
particles that (in contrast to ferromagnets) do not retain a
magnetization upon removing the applied field. This spin degree of
freedom of a magnetic dipole has an obvious parallel with the
internal degree of freedom of molecular motors. Thus are recovered
the features of Maxwell demon-effects, as well as Complexity, that
allows a different route for regeneration in the other half-cycle.
Furthermore, while a monomeric particle (Sect.3.2) in the diffusive
searching phase could even lose track of its initial template, a
dimer of particles would remain associated with the starting
template, if the diffusing phase of the first particle coincided
with the template binding phase of the other.

\section*{3.4 Other motor aspects; magnetocaloric-effect}
Further, in bio-molecular motors no overall macroscopic potential
gradients are present, even if ATP hydrolysis in each cycle has the
effect of raising the local temperature [40]. This 'heat-engine
effect' upon ATP-coupling ensues from vibrational energy drilled
into the molecule. Again, in the entropy-feeding motor mechanism
proposed by Matsuno and Paton [41], coupling to ATP-hydrolysis leads
to release of energy in very tiny quanta (similar in energy to
Brownian hits) and thus an effective temperature of almost zero
Kelvin for the actinomyosin complex. For correspondence to the S-PP
scenario, it is interesting that a direct non-biological mechanism
enabling interchange between a system's environmental temperature
and its own entropy is provided by the (anistropic) magnetocaloric
effect (MCE) [42], which is the property of some magnetic materials
to heat up when placed in an H-field and cool down when they are
removed (adiabatic). And, recent evidence shows that at the
nanoscale too, the heat capacity turns out to be a few-fold higher
than that of bulk systems, thanks to MCE [43]. This effect can also
be seen in another related study, although the paradoxical
phenomenon of cooling by isentropic magnetization in high fields, in
this six-particle system [44], is about one order of magnitude
higher than the conventional cooling mechanism by isentropic
demagnetization and is related to the ring-chain transition. The
cross-over of states emanating from a conflict between magnetic and
structural order underlies this paradoxical effect. In contrast, our
present context involves no such conflict, since it embodies a
conventional magnetic system where diffusing 'hard' S-PPs interact
with H-fields of template partners. Anyhow, this simulation does
provide a concrete example of a nano-scale manifestation of the MCE,
in contrast to ferrofluid systems with a large number of particles
considered for use in magnetocaloric heat engines [44]. And, a
periodic manifestation of MCE -due to two entropic degrees of
freedom (magnetic ($S_M$) vs thermal ($S_T$)) -- follows as a
logical consequence of periodic change superimposed on a background
potential, provided by H-fields of template-partners for S-PPs
drifting along the rock field gradient.

\begin{table} \caption{\textbf{\textbf{Biomolecular Motors vs Particles in MSP
(Hadean)}}}
\begin{tabular}{p{5cm}p{5cm}p{5cm}}
\hline Correspondence in features  & Bio-molecular motor       &
Particle
diffusing through MSP (Hadean) \\
Templates                   & Biopoly-mers like protein filaments
and nucleic acids                                      & Layers of
magnetic particles with dipolar interactions \\
Low Reynolds number         & Nano-particle diffusing in
intracellular viscous mileu                             &
Nano-particle diffusing in magneto-viscous medium \\
Movement direction          & Amino end to Carboxyl end of template
or vice versa                                           & North to
South pole or vice versa; gentle H-field gradient cuts through MSP.Cause vs effect \\
Conversion to mechanical energy directly : Symmetry breaking via
infinitesimal decreased potential in progress direction         &
Gentle decrease in chemical potential upon ATP coupling for
conformations having greater affinity for sites in preferred
direction.                                             & Gentle
decrease in magnetic potential due to gradient, for diffusion in
preferred direction till captured by H-field of template partner. \\
Switching between low and high template-affinity states by
non-eqforce; energy flow from bath        & Via different
conformations due to altered landscape potential by ATP/ADP+Pi
binding/release cycles, where thermal diffusion is rectified & Via
different spin ordering due to altered magnetic potential by
periodic
presence/absence of template partner, helped by thermal excitation from bath \\
Small systems allowing for time-
reversed trajectories             & Nano-scale      & Nano-scale   \\
Time-reversible degree of freedom $\sim$ thermal hits ($k_BT$) & Can
undergo infinitesimal conformational changes.       & Can
undergo infinitesimal spin alignment changes \\
Bond for stabilizing eq state?    & Weak, e.g.
H-bond, etc                                         & Weak -alignment to partner \\
Increased apparent local temp of medium in each cycle and lowered
temp. of that of motor/ magnetic particle?     &Yes, ATP-coupling
leads to enhanced vibrational motion (cannot account for transport
via thermal ratchet); energy released in bits $\sim k_B T$, leading
to nearly 0 K of motor.    & Yes, MCE due to local partner could
cause heat increase of S-PP consequently released to bath, with
spin-relaxation away from partner,
lowering S-PP temp. \\
Processivity vs attachment points  & Enhanced for two heads vs one &
Similar enhancement for dimers vs monomers \\
Self-propelled, template-interactive  transport : non-eq energy and
asymmetry            & Non-eq, time-dependent repetition of
asymmetry ('seen by complex motor') enables generation of drift
velocity by averaging over thermal noise.            & Non-eq,
asymmetry from H-field grad for magnetic dipole. Cyclic template
interactions due to H-fields of template partners in MSP. \\
\hline
\end{tabular}
\end{table}

\section*{3.5 Motor vs magnetic-dipole}
The Table offers some non-trivial parallels between motors moving on
bio-polymers and motion of S-PPs on layers of particles bound by
magnetic dipolar forces. Their nano-size would give a negligible
inertial term in the Langevin equation, which together with
frictional forces, due to viscous medium gives a low Reynold's
number. At each point velocity is the direct result of an external
force, acting on the particle that achieves its terminal velocity
instantaneously; thus thermal hits cause random diffusion. Both
systems offer high efficiency mechanisms for direct conversion of a
non-equilibrium source into mechanical energy, rather than via an
intermediary state, e.g., heat for thermal engines [45].

The slight decrease in magnetic potential energy ( $ \sim  -M
(dH/dz) z $; assuming a constant gradient at small distances, where
M is magnetization) of the diffusing particle has a parallel in the
slight decrease in chemical potential of the motor believed to occur
in the preferred direction of diffusion [5, 34]. Both are examples
of small systems where time-reversed microscopic equations of motion
allow for time-reversed trajectories. In both, thermal hits get
rectified for the periodic recycling between higher and lower
template-affinity states, by close-to-equilibrium coupling. In the
motor, a slow modulation of chemical potential by thermodynamic
energy from ATP hydrolysis biases conformational changes towards
'stickiness' for forward binding sites, on the locally asymmetric
but periodic template. ATP coupling breaks the microscopic
reversibility and drives directed diffusion from N- to C-terminal or
vice versa, with motors binding in similar orientations in either
situation and not facing opposite directions. This asymmetric
template-affinity is remarkably similar to how a gentle increase of
field lines, to the front of or behind, a North to South oriented
dipole can cause its drift in the forward or backward directions,
respectively. Here detailed balance is broken by gentle changes in
flux lines (non-eq) due to rocks, while interactive cycles with
alterations in magnetic ordering are brought about thanks to
alignment with local H-fields of consecutive template-partners in
the MSP. For the ratcheting motor, the different trajectories in the
two half-cycles enable net movement via asymmetric track-binding of
intermediate states. Since trajectory in the first half-cycle is not
retraced, neither is the motor velocity, as template binding
capacity is changed [4, 46, 47]. Herein lies the difference: The
non-equilibrium force does not push the motor directly but by
rectified thermal diffusion via asymmetric motor template
interactions. In contrast, for the diffusing magnetic dipole this
symmetry-breaking drift would have been a direct consequence of the
gentle flux changes, but superimposed local secondary H-fields would
have generated periodic particle-template interactions, with altered
magnetic ordering.

\section*{4.1 Ratchets replaced magnetic effects?}
Clearly, bio-molecular motors are not driven by a macroscopic
external force. But compare this to a slow directed diffusion of a
magnetic dipole in a very gentle gradient due to a non-homogeneous
magnetic field from rocks, a logical scenario rooted in basic
physical principles. To recapitulate, the combination for
self-generated transport-- non-equilibrium and asymmetry -- are both
provided by an H-field for a magnetic particle only; not ordinary
matter. And, template interactions, with cycles between low and high
affinity states, are seen as a consequence of local H-fields of
consecutive partners in the MSP (itself another ramification of the
rock H-field via magnetic-dipolar interactions between the
particles). As to their origins, the first part of the puzzle seems
to be one of searching for a driving force that could have enabled
self-assembly, while simultaneously driving other responses, such as
movement. And, the second is to look for $both$ an external driving
force as well as the complexity of matter being driven. It does not
seem to help if we only search an external force, e.g. a thermal
gradient could have facilitated transport of ordinary matter, but
quickly activated the crystal formation phase, thus limiting access
to a soft self-assembled phase. The other possibility is to look for
matter that could have been present in the Hadean with access to
internal degrees of freedom, underlying the complexity of today's
biomolecules. Conceptually, these could have undergone
self-assembly, and additionally used gradients, e.g. thermal, for
eliciting a response. But how could the emergence of such matter be
explained, in a limited time-frame? Now, in the origins of Life,
S-PPs could have themselves turned into magnetic sources of energy
in the presence of a moderate inducing H-field. It is this
$handshake$ between the magnetic features of a moderate field, S-PPs
(with spin degrees of freedom), and the components of the MSP, that
makes it all conceptually feasible in a Hadean Ocean Floor setting.
It is possible that evolution of this magnetic system translated and
merged the directed gradient-driven diffusion and (interactive)
periodic alterations in magnetic alignment potential therein, into
another with periodic alterations in landscape potential, as in
bio-molecular motors. In the latter, the continuous interactive
movement is via close co-ordination of chemical (hydrolysis) and
mechanical (association-dissociation) cycles. Such evolving
complexity (conformational changes connecting intermediate states
via different trajectories) enabling close-to-equilibrium coupling
to drive the macroscopic system uphill in its landscape potential--a
bio-ratchet-- could have helped disengage the local 'magnetic
ladder'.

\section*{4.2 Search for field controlled assembly in the Hadean}
The search for super-paramagnetic matter that could have been
externally controlled by means of a magnetic-field developed through
serpentinization of the igneous rocks comprising the ocean floor,
led us to the mound scenario conceived by Russell and coworkers [9].
The substance could well have been greigite (a non-stoichiometric
Ni-bearing iron sulphide phase, $\sim$ NiFe$_{5}$S$_{8}$) whose
similarity to complexes in enzymes considered ancient, helped link
Life's origins to the Hadean Ocean Floor.

\subsection*{4.2.1 The mound scenario}
The mound builds up slowly as iron-nickel sulphides precipitate
along with other components in an envisaged environment enriched
with gradients (moderate temperature, redox, pH), leading logically
to a host of metabolites concentrated in membranous compartments,
thereby endowing this scenario with rich metabolic potential (see
Figure 2). Namely, water percolating down through cracks in the hot
ocean crust reacted exothermically with ferrous iron minerals, and
returned in convective updrafts infused with H$_{2}$,  NH$_{3}$,
HCOO$^{-}$, HS$^{-}$, CH$_3^{-}$; this fluid (pH $\sim$ 10 $\leq$
120$^{\circ}$ C), exhaled into CO$_{2}$, Fe$^{2+}$ bearing ocean
waters (pH $\sim$ 5.5 $\leq$ 20$^{\circ}$ C) [48]. The interface
evolved gradually from a colloidal FeS barrier to a single membrane
and thence to more precipitating barriers of FeS gel membranes.
Since fluids in alkaline hydrothermal environments contain very
little hydrogen sulphide, the entry of bisulphide, likely to have
been carried in alkaline solution on occasions where the solution
met sulphides at depth [49], was controlled. This was perhaps
important for the envisaged gel-environment, since colloids often
form more readily in dilute solutions -- suspension as a sol-- than
in concentrated ones where heavy precipitates are likely to form
[28]. Further, theoretical studies by Russell and Hall [50] show the
potential of the alkaline hydrothermal solution (expected to flow
for at least 30,000 years) for dissolving sulfhydryl ions from
sulfides in the ocean crust. The reaction of these with ferrous iron
in the acidulous Hadean ocean (derived from very hot springs [50])
is seen as having drawn a secondary ocean current with the Fe$^{2+}$
toward the alkaline spring as a result of entrainment [51].
Significantly enough, the super-paramagnetic property of greigite
($\leq$ 30-50 nm [52]), one of the components of the FeS colloidal
barrier, brings to light its possible magnetically reproducing
aspect. And, framboids observed in the chimneys [53, plate. 2],
reveal the role of physical forces in producing these dynamically
ordered forms under mound conditions [28].

\begin{figure}
\begin{center}
\includegraphics[width=120mm,height=100mm]{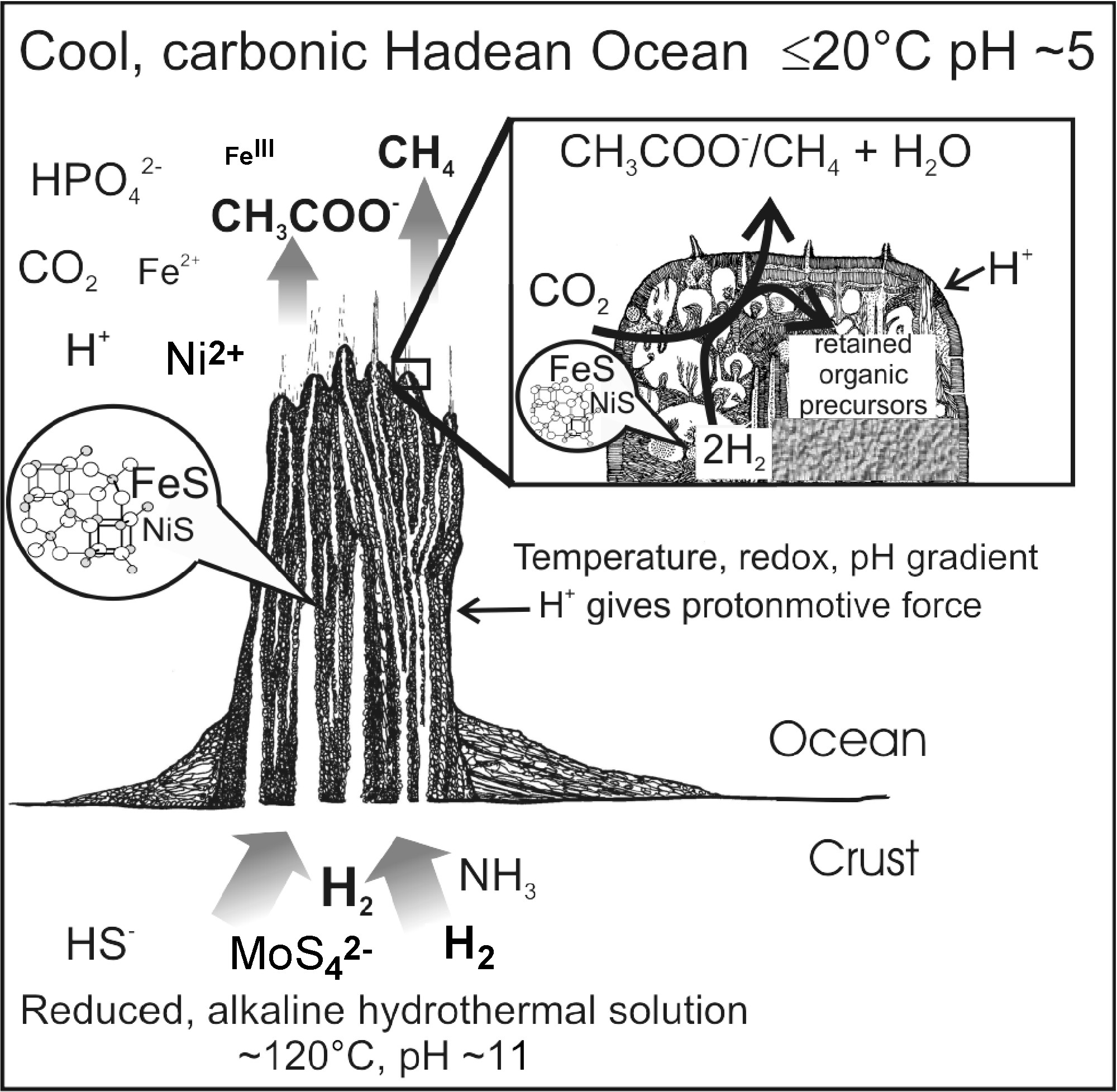} \\
\caption{The hydrothermal mound as an acetate and methane generator.
Steep physicochemical gradients are focused at the margin of the
mound (see text for details). The inset (cross section of the
surface) illustrates the sites where anionic organic molecules are
produced, constrained, react, and automatically organize to emerge
as protolife (from Russell and Martin [54], and Russell and Hall
[49], with permission). Compartmental pore space may have been
partially filled with rapidly precipitated dendrites. The walls to
the pores comprised nanocrystals of iron compounds, chiefly of FeS
[55] but including greigite, vivianite, and green rust occupying a
silicate matrix. Tapping the ambient protonmotive force the pores
and bubbles acted as catalytic culture chambers for organic
synthesis, open to H$_{2}$,  NH$_{3}$, CH$_3^{-}$ at their base,
selectively permeable and semi-conducting at their upper surface.
The font size of the chemical symbols gives a qualitative indication
of the concentration of the reactants.}
\end{center}
\end{figure}

\begin{figure}
\begin{center}
\includegraphics[width=100mm,height=90mm]{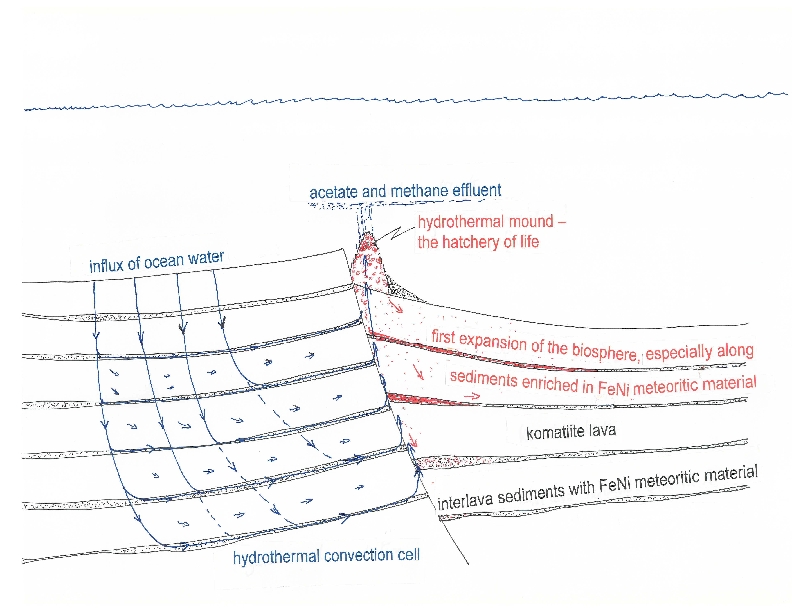}
\caption{Iron and FeNi particles derived from the subhypervelocity
flux of iron and of NiFe-metal-containing chondritic meteorites and
micrometeorites are distributed on the ocean floor below, and
sparsely scattered throughout, the mound. Some of the nickel may
have been incorporated into catalytic mineral clusters such as
greigite; clusters later co-opted into proto-metalloenzymes [50,
13]. Unpublished work of Ostro and Russell (2008), kindly provided
by M.J.Russell.}

\end{center}
\end{figure}

\subsection*{4.2.2 Field estimate from W-B model}
      The associated H-field with rocks, needed for overcoming temperatures $\sim$ 50C in the mound, is estimated by
extrapolating the Wilkin and Barnes (W-B) model [57] for formation
of framboidal pyrite. This is based on the alignment of precursor
greigite (taken as single domain crystals), under the influence of
the weak geo-magnetic field that would help overcome the thermal
energy of particles above a critical size. Ferrimagnetic greigite
has a saturation magnetization value $M_{sat}$ at 298K ranging
between 110 and 130 $kA/m$. Assuming a spherical geometry, the
critical grain diameter of constituent crystallites comprising the
framboid interior $d_c = 2a$, where $a > 1$, is given by
\begin{equation}\label{4.1}
d_c = (6 k_BT /\mu_0 \pi M_{sat}|H|)^{1/3}
\end{equation}
This result can be obtained from the inequality $ W_{WB} > k_B T$
where we $define$ $ W_{WB} \equiv \mu_0 M_{sat} V H $. Here $k_B$ is
the Boltzmann's constant and $\mu_0$ the permeability of vacuum.
When aligned parallel to weak geomagnetic field ($\sim$ 70$\mu$T),
$d_c$ = 0.1 $\mu$m. According to this formula, a rock H-field for
accreting 10nm sized particles would have to be 1000 fold higher.
This also is of the same order of magnitude $\sim$ 10mT, seen for
magnetite-based ferrofluids [16]. For, the saturation magnetization
of magnetite ($M_s$ = $4.46 \times 10^5 $ A/m) is about 3.5 times
greater than that of greigite; from this one expects proportionate
values for the fluid susceptibility of a corresponding greigite
suspension, building up slowly in the ocean waters (see above).
Further, the dipole-dipole interactions between negatively charged
greigite particles, under mound conditions where pH is well above 3
[57] is likely to be aided by the screening effect due to ionic
strength of natural waters [29]. This description of an aqueous
suspension of super-paramagnetic greigite matches that of aqueous
ferrofluids, although the particles would lack the protective
(steric-stabilization) coating of their synthetic counter-parts.
Hence such a dispersion should show non-ideal behaviour even under
dilute conditions, where magnetic dipole forces would attract
particles, as in non-ideal ferrofluids where dissipative internal
structures are known to form in the presence of an external field.
Of course, if energy due to dissipative structure formation were
also included, the effective local field would be only higher ; this
can be checked by comparing $W_{WB}$ (see above) with the equation
$W_T = W_M + W_S$ of Li et al [21] for a non-ideal ferrofluid (Sect
2.1), where the (positive) second term $W_S$ represents the effect
of interaction, and effectively increases the value of the H-field
present in the first term $W_M$.

\par
Thus a moderate rock H-field $\sim$ 50 -100mT would lead to magnetic
accretion of nano-sized greigite particles -- a soft magnetic
assembly (a new phase) -- $not$ accessible without an H-field.
Magnetic dipolar forces would provide the compression for `packing'
(in 3d-space) in this finite system, giving it access to aperiodic
order. Indeed, this very logic underlies the `magnetic reproduction
proposal' of Breivik [58] where his model of templates formed from
ferromagnetic `monomers' offers a means to study the direct link
between thermodynamic and the information-theoretic concept of
entropy.

\subsection*{4.2.3 Local field due to magnetic rocks}
So the issue is to look for how magnetic rocks could provide a
moderate local H-field $\sim$ 50-100mT. Now, low levels of
magnetization in rocks leading to crustal magnetic anomalies on the
present day Ocean floor are typically understood in terms of (apart
from mechanisms like sedimentation) the classical mechanism of
thermo-remnant magnetization (TRM)  -- acquired when newly formed
minerals cool below their Curie temperature in the presence of the
geo-magnetic global field. On the other hand, for achieving a local
field we note that subsurface magnetic rocks are known to create
sufficiently intense magnetic anomalies (w.r.t. geo-magnetic field)
used to track their location. As an example consider the rich iron
ore province in the Pilbara region of Western Australia with a
background ambient magnetic field of about 55 $\mu T$, where a
helicopter survey recorded high anomaly amplitudes of up to 120 $\mu
T$, indicating the high percentage of iron ore composition [59].
Since field strength decreases rapidly with distance ($\sim r^{-3}$)
from the magnetic medium, the corresponding value on the rock
surface is expected to be higher by a few orders of magnitude. This
overwhelms the contribution of the ambient geomagnetic field, which
was already about half as strong 3.2 billion years ago as it is
today [60]. A stronger reason for the irrelevance of the
geo-magnetic field vis-a-vis local (rock) H-fields comes from the
fact that $\sim$ 4.1-4.2 Ga, the time when Life is believed to have
been already initiated ($\sim$ 4.2-4.3 Ga [9, 50]), the geomagnetic
field did not even exist (!) [61]. This leaves the local field due
to magnetic rocks as a primary candidate governing the initial
conditions leading to Life.

\subsection*{4.2.4 Magnetism from extraterrestrial sources}

Now the present geomagnetic field strength is too weak to explain
the high NRM (natural remanent magnetization) to SIRM (saturation
isothermal remanent magnetization) ratios of lodestones, i.e.,
natural magnets with magnetic field strengths varying upto 0.1 Tesla
[see 62], as the initial magnetization depends on the strength of
the inducing field. This eventually led to lightning remnant
magnetism as a plausible mechanism [63]. Further, T\'unyi et al.
[64] have examined the possibility of nebular lightnings as a source
of impulse magnetic fields (in the context of accretion of Earth and
other planets, that is seen as rendering the gravitational accretion
process more efficient) by magnetizing the ferromagnetic dust grains
to their saturation levels. Quite possibly, the most important
contribution to the crust was thanks to the presence of accreted
highly magnetized meteoritic matter, with acquired isothermal
remanent magnetism, such as seen in meteorite, lunar samples, etc.
Indeed, the Wasilewski group [65] propose magnetization of
chondrules that cooled while spinning and translating through a
magnetic field, in view of their matching demagnetization profiles
with that of melt slag droplets. They also employ the properties of
metallic systems for explaining remanence in lunar and meteoritic
samples containing iron and iron alloys in contrast to that of
terrestrial ones comprising oxides. And, they describe specific
structures and microstructures associated with magnetic remanence
effects for the Fe-Ni system, produced by various transitions and
transformations with or without diffusion [66].

\subsection*{4.2.5 Extra-terrestrial magnetic matter in the mound}
Now, the presence of ferromagnetic matter due to vestiges of iron
and iron-alloy-containing meteorite bodies in the primitive Hadean
crust, seems relevant in view of  conditions in the primitive crust
that were highly reducing (in contrast to today's picture) with the
redox state depicted at Fe-FeO (Wustite) [67, 68]. While the oceans
are believed to have been formed around 4.3Ga, life is thought to
have emerged between 4.3 and 4.2 Ga [9, 50], when conditions in the
newly formed crust still seem to have been extremely reducing [61].
Indeed, impact craters formed by asteroids and comets that offer a
route for delivery of extraterrestrial iron from iron-containing
meteorites, have been pointed out as hosting conditions important
for the emergence of Life, e.g. catalytic reduction of $CO_{2}$ that
is linked to the origins of metabolic pathways [69]. In an extension
of this scenario, Ostro and Russell (2008; unpublished results
kindly provided by MJ Russell) suggest that similarly reducing Ocean
floor accumulations may also have resulted from the non-cratering
(sub-hypervelocity) flux of NiFe-metal containing meteorites and
micrometeorites onto the Earth's surface. As shown in Figure 3, in
addition to acetate production by reduction of dissolved $CO_{2}$ by
precipitated Fe(II)-bearing minerals (Figure 2), the presence of Fe
and FeNi particles accumulated around the base of the mound could
have allowed $CO_{2}$ reduction all the way to $CH_{4}$. Their
analysis is based on extrapolating available statistics on current
flux of extraterrestrial matter vis-a-vis its metal fraction back
over four billion years. Although exposure to water is expected to
lead to corrosion, apart from the fact that external oxidized layers
would hamper the weathering process, they argue that the presence of
nickel would have helped in enhancing the resistance of meteoritic
metal to oxidation (as in stainless steel alloys). They have pointed
out that owing to the then powerful tidal currents [70], dense
meteoritic matter -from fine grained particles to larger ones--
would tend to be trapped in local basins, e.g. collecting around
protuberances like hydrothermal mounds. Thus, in such an environment
with possibility of magnetic elements (magnetite, awaruite, and
iron-nickel alloys), as observed in meteorites [61] the chances of
producing a local magnetic environment seems highly plausible. Still
another possibility is an internal mechanism like spontaneous
magnetization.

\subsection*{4.2.6 Reinforcing H-field by serpentization}
In an earlier version we had simply assumed the presence of magnetic
rocks in the mound, say via mechanisms such as lightning remnant
magnetism [66] and on the lines of T\'unyi et al [64] that could
have provided a local field up to $\sim$ Tesla [24(ii)], while only
moderate fields $\sim$ tens of mTesla would suffice for accretion.
Improving on this scenario, magnetic rocks are seen as situated
immediately beneath the mound and to have been produced during the
serpentinization of ocean floor peridotites in a process that
generates magnetite [71, 23], and also awaruite [72].

\par
Further, in the present paper, molecular motor-like diffusion
(close-to-equilibrium) of greigite nano-particles through the MSP is
envisaged as being propelled by a gentle flux gradient - a scenario
which is rather naturally simulated by the non-homogeneous H-field
generated by magnetic rocks. This is in contrast to an earlier
approach [24] where we had considered the possibility of the
temperature gradient in the hydrothermal system itself as having
driven this molecular motor-like passage, and therefore the possible
co-evolution of both reproducing and metabolic aspects of greigite
simultaneously in the same location. But the problem of a
$single-location$ origin, in dealing with far-from-equilibrium
gradients supporting metabolism with close-to-equilibrium driven
diffusion in an identical location (the MSP being a delicate phase),
has necessitated a change: a close but separately situated origins
of the $two$ wings of Life. And, not too far from the gradient,
magnetic rocks in cooler waters, would coax a gentle build up of
greigite particles into an MSP (aided by MCE due to the rock field
that could give rise to a mild turbulence). The complexity of the
MSP, in turn, would evolve thanks to a continuous supply of
chemicals diffusing from the metabolic counterpart in the mound.
It's coupling to energy-rich ones may have led to bio-like ratchets,
permitting exit from the confines of the magnetic rock field, with
the two faces of greigite enabling complex energy transduction
mechanisms (see Sect.4.3).

\section*{4.3 Conclusions: Double-origins revisited}
The vicinity of a physical self-reproducer (via magnetic
rock-controlled S-PPs) to its metabolic counterpart, as in the
proposed double origins [8], could have allowed replacements via a
'chemical genie'. And a driven system where a coherent energy source
- the H-field - maintained phase correlations between constituents
of the assembly would provide a natural selection basis for its
chemical replacements with capacity for such anisotropic dynamics in
the absence of the H-field. This is in contrast to conventional
proposals of randomly evolving chemical reactions that by chance,
led to the emergence of Life but in a non-specified time frame. The
requirements of a {\it starting magnetically controlled phase} do
offer a basis for explaining the emergence of coherently coupled
systems comprising non-equilibrium sources like ATP on the one hand,
and on the other of evolving soft matter with their internal degrees
of freedom that can exist in different equilibrium states,
inter-convertible by harnessing random Brownian motion. And note
that the potential of magnetic particles for evolving transduction
mechanisms lie in their capacity to interact with other sources of
energy. For example, a magnetic Soret effect [73] can provide a
means for rectified diffusion, on analogous lines to the
thermophoretic Soret-effect [74], due to infinitesimal changes in:
susceptibility vs solute-solvent interfacial tension, respectively.

Finally, a remarkable spin-off of directed movement of cargo-loaded
magnetic particles, across a packed array, is a logical
symmetry-breaking enrichment of one from a pair of ligated optical
isomers, by the `grinding effect' [see 75] due to space constraints
on surface-transfer reactions. Note that magnetic effects can
non-invasively resolve intractable mixtures [76] of magnetic and
non-magnetic components; they can also show up invasively by
controlling spin states in biological systems: from chemical
reactivity (due to spin-selectivity of reactions; see [77]) to
quantum coherence [78]. The Maxwell Demon-like potential of S-PPs
diffusing through an MSP, due to a gentle H-field gradient gives
access to coherent dynamics. Indeed, such a system seems to fit the
requirements of Davies [79] quantum computing origins-of-life
proposal, as also acknowledged by him in [80]. Again, the larger
information storage-capacity of a DNA-motor system, than the usual 1
bit/base basis, results from several internal states of the motor
itself [81]. And, such an information-processing network of
DNA-motor-bath could have its natural origin in a $magnetic$
Ancestor.

The common material constituents comprising all kingdoms of Life are
certainly important clues for the origins of Life. But, the list of
commonalities also feature non-material aspects like sense,
induction, search capacity, sensitivity to fields, adaptation
potential, feedback within a hierarchially assembled network where
local units dictated collective properties emerging at the global
level, to name some. The traditionally accepted picture is that
these key features of Life evolved at different space-times but
merged together somehow to produce the replicating wing of Life. The
other possibility considered here is a simple physical system having
the above physical properties, including the capacity for
computational searches. Armed with this potential and with help from
the metabolic arm of the origins of life, it could have set about
'training' chemistry, persuading it to behave like it does in
biology, instead of as in non-living systems. Did we have a
computing Ancestor directing its own evolution? Maybe
condensed-matter physics could help in this search \ldots

Acknowledgements: We thank Prof. M.J. Russell for inspiration and
kind support with figures, data, plus references; the anonymous
Referee for his suggestions and key references; Prof. K. Matsuno for
suggesting a closer look at electrostatic effects; Prof. A.K. Pati
for bringing "Quantum Aspects of Life" to our notice.  This work was
entirely financed, with full infrastructural support, by Dr.
Jean-Jacques Delmotte; Drs A. Bachhawat and B. Sodermark gave gentle
push;  Dr. V. Ghildyal and Mr. Vijay Kumar helped with manuscript
processing.


\begin{thebibliography}{99}

\bibitem{1}
C. Bustamante, J. Liphardt, F. Ritort, Physics Today, 58, 43 (2005)

\bibitem{2}
T. Harada, S. Sasa, Math Biosci. 207(2), 365 (2007)

\bibitem{3}
J. Weber, A.E. Senior, FEBS Letts. 545, 61 (2003)

\bibitem{4}
R.D. Astumian, P. Hanggi, Physics Today, 55(11), 33 (2002)

\bibitem{5}
R.D. Astumian, Proc. Natn. Acad. Sci. 104(50), 19715 (2007)

\bibitem{6}
R.D. Vale, Trends Cell. Biol. 9, M38 (1999)

\bibitem{7}
A.G. Cairns - Smith, Seven clues to the origin of life (Cambridge
University Press, New York, 1985)

\bibitem{8}
F.J. Dyson, Origins of Life, 2nd ed. (Cambridge Univ Press,
Cambridge 1999)

\bibitem{9}
M.J. Russell, A.J. Hall, J. Geol. Soc. London 154 (pt 3) 377 (1997)

\bibitem{10}
J. Reitner, J. Peckmann, A. Reimer, G. Schumann, V. Thiel, Facies
51, 66 (2005)

\bibitem{11}
S.L. Simmons, D.A. Bazylinski, K.J. Edwards, Science 311, 371 (2006)

\bibitem{12}
J.L Kirschvink, J.W. Hagadorn, In : E. B\"auerlein (Ed.) The
Biomineralization of Nano- and Micro- structures (Wiley VCH,
Weinheim, Germany, 2000) 139

\bibitem{13}
E.J. Milner-White, M.J. Russell, Origins Life Evol. Biosphere, 35,
19 (2005)

\bibitem{14}
A.G. Cairns-Smith, Chem. Eur. J., 14, 3830  (2008)

\bibitem{15}
E. Schr\"odinger,  What is life? The physical aspects of the living
cell (Cambridge University Press, Cambridge, 1944)

\bibitem{16}
S. Odenbach, J. Phys. Condens. Matter 16, R1135 (2004)

\bibitem{17}
A.Y. Zubarev, L.Y. Iskakova, Physica A 343, 65 (2004)

\bibitem{18}
A.Y. Zubarev, J. Fleischer, S. Odenbach, Physica A 358, 475 (2005)

\bibitem{19}
Z. Wang, C. Holm, Phys. Rev. E 68(4), 041401 (2003)

\bibitem{20}
S. Taketomi, H. Takahashi, N. Inaba, H. Miyajima, J. Phys. Soc. Jpn.
60 (5), 1689 (1991)

\bibitem{21}
J. Li, Y. Huang, X. Liu, Y. Lin, L. Bai, Q. Li, Sci. Tech. Adv.
Mater. 8, 448 (2007)

\bibitem{22}
R. Pastor-Satorras, J.M. Rubi, J. Magn. Magn. Mater. 221(1-2), 124
(2000)

\bibitem{23}
T. Schroeder, B. John, R. Frost, Geology; 30(4), 367 (2002)

\bibitem{24}
G. Mitra-Delmotte, A.N. Mitra, (i) arXiv:0710.0220v1 [cond-mat.soft]
(2007); (ii) arXiv:0809.3316v3 [cond-mat.soft]

\bibitem{25}
R.E. Rosensweig, Ferrohydrodynamics (Dover, New York, 1997)

\bibitem{26}
B. Huke, M. L\"ucke, Rep. Prog. Phys. 67 (10), 1731 (2004)

\bibitem{27}
C. Timm, Phys. Rev. E 66, 011703 (2002)

\bibitem{28}
Z. Sawlowicz, Prace Mineralog. Pan. 88, 1 (2000)

\bibitem{29}
J. Spitzer, B. Poolman, Microbiol. Molec. Biol. Reviews, 73, 371
(2009)

\bibitem{30}
J.T. Trevors, G.H. Pollack, Prog. Biophys. Mol. Biol. 89(1), 1
(2005)

\bibitem{31}
A. Vologodoskii, Phys. Life Rev. 3, 119 (2006)

\bibitem{32}
Y. Taniguchi, M. Nishiyama, Y. Ishii, T. Yanagida, Nat. Chem. Biol.
1(6), 319 (2005)

\bibitem{33}
S. Duhr, D. Braun, Phys. Rev. Lett. 96, 168301 (2006)

\bibitem{34}
R.D. Astumian, Proc. Natn. Acad. Sci. 104(1), 3 (2007)

\bibitem{35}
C.R. Vestal, Ph.D. Thesis, Georgia Institute of Technology (Atlanta,
GA, USA, 2004)

\bibitem{36}
A. Engel, P. Reimann, Phys. Rev. E 70, 051107 (2004)

\bibitem{37}
V. Becker, A. Engel, Physica A 354, 59 (2005).

\bibitem{38}
S. Ban, V. Korenivski, J. Appl. Phys. 99, 08R907 (2006)

\bibitem{39}
R. Palm, V. Korenivski, New J. Phys., 11, 023003 (2009)

\bibitem{40}
P. Reimann, P. Hanggi, Appl. Phys. A 75, 169 (2002)

\bibitem{41}
K. Matsuno, R.C. Paton, Biosystems 55(1-3), 39 (2000)

\bibitem{42}
A.M. Tishin, Y.I. Spichkin, The Magnetocaloric Effect and its
Applications (IOP Publishing, Bristol and Philadelphia, 2003) 475

\bibitem{43}
V.V. Korolev, I.M. Arefyev, A.G. Ramazanova, J. Thermal Anal. Cal.,
92(3), 691 (2008)

\bibitem{44}
P. Borrmann, H. Stamerjohanns, E.R. Hilf, D. Tomanek, Eur. Phys. J.
B 19, 117 (2001)

\bibitem{45}
F. Fulga, D.V. Nicolau Jr., D.V. Nicolau, Integr. Biol., 1, 150
(2009)

\bibitem{46}
S. Dasmahapatra, J. Werner, K.-P. Zauner, Int. J. Unconventional
Computing, 2, 305 (2006)

\bibitem{47}
R.A. Cross, N.J. Carter, Current Biol. 10 R177 (2000)

\bibitem{48}
M.J. Russell, N.T. Arndt, Biogeosciences 2, 97 (2005)

\bibitem{49}
M.J. Russell, A.J. Hall, In: L. Zaikowski and J. M. Friedrich (Eds.)
"Chemical Evolution II: From Origins of Life to Modern Society".
(American Chemical Society, 2009, in press)

\bibitem{50}
M.J. Russell, A.J. Hall, In: S.E. Kesler, H. Ohmoto (Eds.) Evolution
of early earth's atmosphere, hydrosphere, and biosphere-constraints
from ore deposits. Geological Society of America, Memoir, 198, 1
(2006)

\bibitem{51}
W. Martin, J. Baross, D. Kelley, M.J. Russell, Nature Rev.
Microbiol., doi:10.1038/nrmicro1991 (2008)

\bibitem{52}
V. Hoffmann, Phys. Earth Planet. Interiors 70, 288 (1992)

\bibitem{53}
R.C.L. Larter, A.J. Boyce, M.J. Russell,  Mineralium Deposita, 16,
309 (1981)

\bibitem{54}
M.J. Russell, W. Martin, Trends Biochem. Sci. 29(7), 358 (2004)

\bibitem{55}
M. Wolthers, S.J. Van Der Gaast, D. Rickard, Amer. Mineral. 88, 2007
(2003)

\bibitem{56}
M.J. Russell, A.J. Hall, A.J. Boyce, A.E. Fallick, Econom. Geol. 100
(3), 419 (2005)

\bibitem{57}
R.T. Wilkin, H.L. Barnes, Geochim. Cosmochim. Acta 61, 323 (1997)

\bibitem{58}
J. Breivik, Entropy 3, 273-279 (2001)

\bibitem{59}
Fox Resources Ltd. (ASX Media Announcement: High magnetics confirm
iron ore potential at Mt. Oscar), 17 Oct. 2007;
http://www.foxresources.com.au/documents/318.pdf

\bibitem{60}
J.A. Tarduno, R.D. Cottrell, M.K. Watkeys, D. Bauch, Nature, 446,
657 (2007)

\bibitem{61}
R. Hazen, D. Papineau, W. Bleeker, R. T. Downs, J. M. Ferry, T. J.
McCoy, et al, American Mineralogist, 93, 1693 (2008)

\bibitem{62}
AG Education Services Ltd. (Magnetism and magnetic surveys,
Geological Survey of Ireland), Science and Technology in Action, 3rd
Edition (2007); www.sta.ie

\bibitem{63}
P. Wasilewski, G. Kletetschka, Geophys. Res. Lett., 26(15), 2275
(1999)

\bibitem{64}
I. T\'unyi, P. Guba, L.E. Roth, M. Timko, Earth, Moon, and Planets
93 (1), 65 (2003)

\bibitem{65}
P.J. Wasilewski, T.L. Dickinson, Meteoritics, 30(5), 594 (1995)

\bibitem{66}
P. Wasilewski, Earth, Moon and Planets, 9(3-4), 335 (1974)

\bibitem{67}
K. Righter, M.J. Drake, G. Yaxley, Phys. Earth Planet. Interiors
100, 115 (1997)

\bibitem{68}
M.J. Russell, A.J. Hall, A.R. Mellersh, In: R. Ikan (Ed.) Natural
and laboratory-simulated thermal geochemical processes (Dordrecht,
Kluwer Academic Publishers, 2003) 325

\bibitem{69}
C.S. Cockell, Phil. Trans. R. Soc. B 361(1474), 1845 (2006)

\bibitem{70}
H.-J. Brink, Z. dt. Ges. Geowiss. 157(1), 17 (2006)

\bibitem{71}
J. Dyment, J. Arkani-Hamed, A. Ghods, J. Int. 129, 691 (1997)

\bibitem{72}
J.S. Beard, L. Hopkinson, J. Geophys. Res. 105(B7), 16527 (2000)

\bibitem{73}
Th. V\"olker, E. Blums, S. Odenbach, J. Magn. Magn. Materials, 252,
218 (2002)

\bibitem{74}
D. Braun, A. Libchaber, Phys. Biol. 1, P1 (2004)

\bibitem{75}
J.M. McBride, J.C. Tully, Nature 452, 161 (2008)

\bibitem{76}
A.W. Schwartz, Chem.Biodiv. 4(4), 656 (2007)

\bibitem{77}
A.L. Buchachenko, Pure Appl. Chem., 72 (12),  2243 (2000)

\bibitem{78}
I.K. Kominis, arXiv:0804.3503

\bibitem{79}
P.C.W. Davies, In : D. Abbott, P.C.W. Davies, A.K. Pati (Ed.s)
Quantum aspects of life (Imperial College Press, London,  2008) 3

\bibitem{80}
Z. Merali, New Scient., 196(2633), 6 (2007)

\bibitem{81}
A. Goel, In : D. Abbott, P.C.W. Davies, A.K. Pati (Ed.s) Quantum
aspects of life (Imperial College Press, London, 2008) 97
\end{thebibliography}
\end{document}